\documentclass[twocolumn,prl,aps,showpacs,floatfix]{revtex4}
\usepackage{epsfig}

\def\bk{\mathbf{k}}
\def\kp{\mathbf{k_\parallel}}
\def\Gbar{\overline{\Gamma}}

\begin{document}

\title{Ballistic Spin Injection from Fe(001) into ZnSe and GaAs}

\author{O. Wunnicke}
\author{Ph. Mavropoulos}
\author{R. Zeller}
\author{P.H. Dederichs}
\affiliation{Institut f\"ur Festk\"orperforschung, Forschungszentrum
J\"ulich, D-52425~J\"ulich, Germany}
\author{D. Grundler}
\affiliation{Institut f\"ur Angewandte Physik und Zentrum f\"ur
  Mikrostrukturforschung, Universit\"at Hamburg, Jungiusstra\ss e 11,
  D-20355 Hamburg, Germany}

\date{\today}

% Abstract ----------------------------------------
\begin{abstract}

We consider the spin injection from Fe into ZnSe and GaAs in the
ballistic limit. By means of the \textit{ab initio} SKKR method we
calculate the ground state properties of epitaxial Fe$|$ZnSe(001) and
Fe$|$GaAs(001) heterostructures. Three injection processes are
considered: injection of hot electrons and injection of ``thermal''
electrons with and without an interface barrier. The calculation of
the conductance by the Landauer formula shows, that these interfaces act
like a nearly ideal spin filter, with spin polarization 
as high as 99\%. This can be traced back to the symmetry of the band
structure of Fe for normal incidence.

\end{abstract}

% 72.25.Hg :  Electrical injection of spin polarized carriers
% 72.25.Mk :  Spin transport through interfaces
% 73.23.Ad :  Ballistic Transport

\maketitle

% Text --------------------------------------------
The controlled injection of a spin polarized current into a
semiconductor (SC) is one of the central problems in the new field of
spin electronics, since it is a prerequisite for the development of
new spin dependent devices \cite{Datta}. Recently some important
successes have 
been achieved. Fiederling \textit{et al.} \cite{Fiederling} have
demonstrated the 
injection from the paramagnetic II-VI SC
Be$_x$Mn$_y$Zn$_{1-x-y}$Se into GaAs with very high spin polarization
using an external magnetic field, while Ohno \textit{et al.}
\cite{Ohno} were able 
to show the injection from the ferromagnetic SC
Ga$_{1-x}$Mn$_{x}$As into GaAs with an efficiency of
1\%. However both methods have the disadvantage that they require a
low temperature. Therefore the injection from a ferromagnet with
large Curie temperature such as Fe would have strong
advantages. Such attempts, though, have not been very
successful in the past, i.e., the reported spin injection efficiency
was low \cite{Hammar,Filip}. Schmidt \textit{et al.}
\cite{Schmidt} 
revealed that a basic obstacle for spin injection from
a ferromagnetic metal into a SC exists, being represented
by
the large conductivity mismatch between both materials. Nevertheless
Rashba \cite{Rashba} as well as Fert and Jaffr\`es \cite{Fert} have
recently pointed out, that this obstacle can be overcome by
introducing a tunneling barrier. 
Meanwhile, and independently, Zhu  \textit{et al.} \cite{Zhu} were
successful in demonstrating the spin injection 
at room-temperature from Fe(001) into GaAs with an
efficiency of 2\% which they attributed to tunneling through a
Schottky barrier.

Kirczenow \cite{Kirczenow} has lately pointed out that contrary to
the ferromagnet$|$metal interface the interface between a ferromagnet
and a SC could act as an ideal spin filter, if e.g. the
Fermi surface of the majority or minority spin bands has a hole at the
$\Gbar$-point of the two-dimensional Brillouin zone, so that
only electrons of the other spin band can scatter into the conduction
band states of the SC at the $\Gbar$-point. However
relevant hybrid systems like Fe$|$ZnSe(001) and Fe$|$GaAs(001) for
which epitaxial growth has been demonstrated, do not show this simple
property.

Recently two ballistic calculations \cite{Grundler, Hu} for
the spin injection process have been published, which basically rely
on a free-electron description of the majority and minority
spin bands. Grundler \cite{Grundler} could argue in this way that the
Fe$|$SC interface can act as a spin filter with an
efficiency of a few percent. Motivated by this work we present here an
\textit{ab initio}
calculation of the ground state properties and the ballistic transport
through the Fe$|$ZnSe(001) and Fe$|$GaAs(001) interfaces. 
In contrast to the above mentioned methods our calculations
include the whole complexity of the band structures of the ferromagnet
and the SCs as well as the even more complex properties of
the interface. The important result of our calculation is, that the
considered Fe$|$SC interfaces act like nearly ideal
spin filters, with spin injection ratios as high as 99\%. We can
attribute this to the 
different symmetries of the majority and minority $d$-bands of Fe at
the Fermi level, a behavior which cannot be described in the
free-electron model. Taken
together with the results of Zhu \textit{et al.} \cite{Zhu} our
calculations give a bright outlook for the spin injection from
ferromagnetic Fe into SCs. 

Our method is based on the local density approximation of
density functional theory and apply the screened KKR-method
\cite{Wildberger}. The heterostructure consists of a Fe halfspace and a
SC (either ZnSe or GaAs) halfspace, both oriented in
the (001) direction and being epitaxially bonded at the interface, so
that the SC lattice constant is double the Fe constant
(a$_\mathrm{Fe}^\mathrm{exp}$ = 5.425~a.u. is used in the
calculation). The two halfspace Green's functions are determined by
the decimation technique \cite{Turek}. In the interface region the
potentials of 4 monolayers (ML) of Fe and 2 ML of SC are
determined selfconsistently. The potentials of all other ML are
identified with the asymptotic bulk values. In all calculations we use
a cut-off of $\ell_{\mathrm{max}} = 2$ for the wavefunctions and an
atomic-sphere-approximation for the potentials, but include the full
charge density. The ballistic conductance $G$ is calculated by the
Landauer-B\"uttiker formalism for $T = 0$. Here we use an expression
similar to the 
one derived by Baranger and Stone \cite{Baranger}, but adjusted to the
asymptotic Bloch character of the wavefunctions and the
two-dimensional translation symmetry of the system. The in-plane
component $\kp$ of the $\bk$-vector enumerates then the scattering
channels, and we can express the $\kp$-dependent conductance $G(\kp)$
wholly in terms of the Green's function of the system. Spin-orbit
coupling is neglected in the calculation.

\begin{figure} [tb]
  \begin{center}
    \epsfig{file=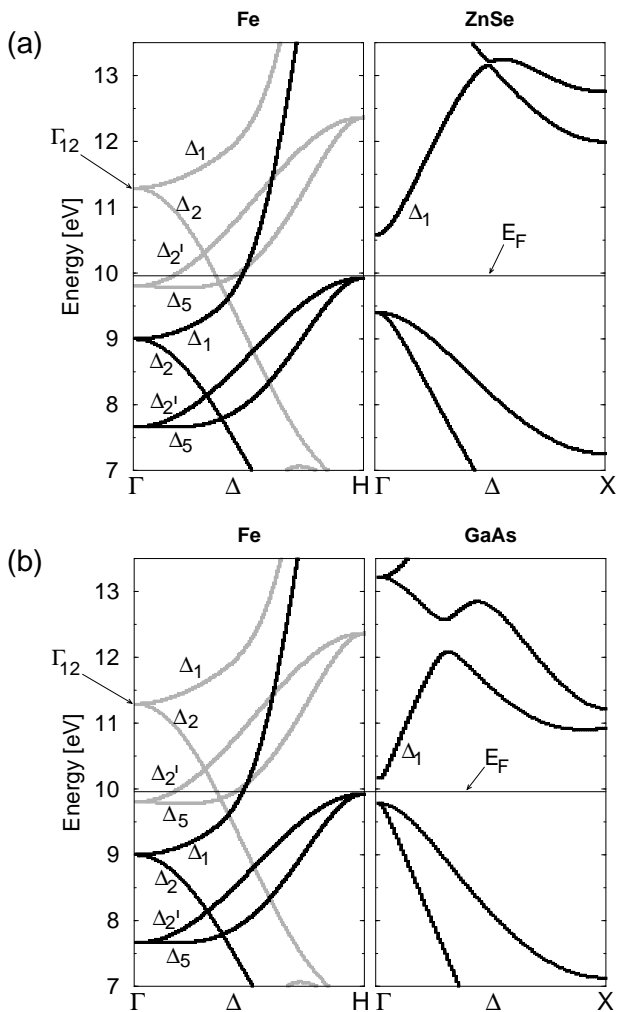,width=3.2in,angle=0}
    \caption{Band structure of Fe (left panel) and the semiconductor
      (right panel): ZnSe (a) and GaAs (b)
      around the Fermi energy. The black lines in the Fe band
      structure are the majority and the gray lines the minority spin
      bands. The $\bk$-vector is varied along the (001)-direction
      $\Delta$.}
    \label{fig1}
  \end{center}
\end{figure}   
As we will demonstrate in this paper, the spin injection process is to
a large extent determined by the symmetries of the bulk
band structures. For this reason we show in Fig.\ \ref{fig1} the
band structure of Fe and of the SCs ZnSe (a) and GaAs (b),
for Bloch vectors $\bk = (0,0,k_z)$ normal to the interface. 
These are the
states relevant for the injection process, since in the SC only states
close to the conduction band minimum $E_C$ will be populated, having 
$\kp \approx 0$.
The left panel shows the spin split majority and
minority bands of Fe in the region of the Fermi level $E_F$.
As usual the different bands in (001) direction are
indexed by $\Delta_1, \Delta_2$, etc. indicating the symmetries of the
wavefunctions \cite{Callaway}. On the right side, the SC bands are
shown with $E_F$ assumed to be located in the middle of the gap. Most
important is here that the lowest conduction states have
$\Delta_1^\mathrm{SC}$-symmetry; they are invariant under all
symmetry operations of the zink-blende lattice, that transform the
Bloch vector $\bk = (0,0,k_z)$ in itself. These operations form the
symmetry group $C_{2v}$, which is at the same time identical with the
symmetry of the whole Fe$|$SC(001) interface. It is now important to
single out those Fe states, which are compatible with this $C_{2v}$
symmetry. In Fe, the $\Delta$-nomenclature refers to the $C_{4v}$
symmetry group, since, contrary to the zink-blende lattice, in the bcc
lattice the (001) direction is a fourfold axis. Thus, not only the
$\Delta_1^\mathrm{Fe}$-states, consisting locally of $s, p_z$ and
$d_{z^2}$ orbitals, can couple to the $\Delta_1^\mathrm{SC}$-band
states, but also the $\Delta_{2'}^\mathrm{Fe}$-states consisting
locally of in-plane $d_{xy}$ orbitals. Here we assume that the $x$
and $y$ directions point along the cubic axes. On the other hand the
Fe states of $\Delta_2^\mathrm{Fe}$-symmetry (with $d_{x^2-y^2}$
character) as well as the Fe states with
$\Delta_5^\mathrm{Fe}$-symmetry (with $p_x$ and $d_{xz}$ or $p_y$
and $d_{yz}$ character) cannot couple to the
$\Delta_1^\mathrm{SC}$-states, since they do not show the full
symmetry $C_{2v}$ of the 
heterostructure. For the spin injection it is now important, that in
the majority band at $E_F$ and above there exists only a
$\Delta_1^\mathrm{Fe}$-band (below $E_F$ also a $\Delta_{2'}$-band is
available) while in the minority band around $E_F$ only a
$\Delta_{2'}^\mathrm{Fe}$-band exists that can couple to the
$\Delta_1^\mathrm{SC}$-states, since the
$\Delta_1^\mathrm{Fe}$-band appears here at about 1.3~eV above $E_F$
(see $\Gamma_{12}$ in Fig.\ \ref{fig1}).

Not shown in Fig.\ \ref{fig1} is the lower $\Delta_1^\mathrm{Fe}$-band 
separated from the upper $\Delta_1^\mathrm{Fe}$-band by the so-called
$s$-$d$ hybridization gap. This gap is characteristic for the
transition metals and arises from the hybridization of the $s$ with
the $d_{z^2}$ orbitals. For the (001) orientation this gap is so
large that for the minority band $E_F$ lies in the gap, giving rise to
the spin filtering effect discussed in this paper. This effect
is also important in magnetic tunnel junctions
\cite{Mavropoulos}. 

\begin{figure} [tb]
  \begin{center}
    \epsfig{file=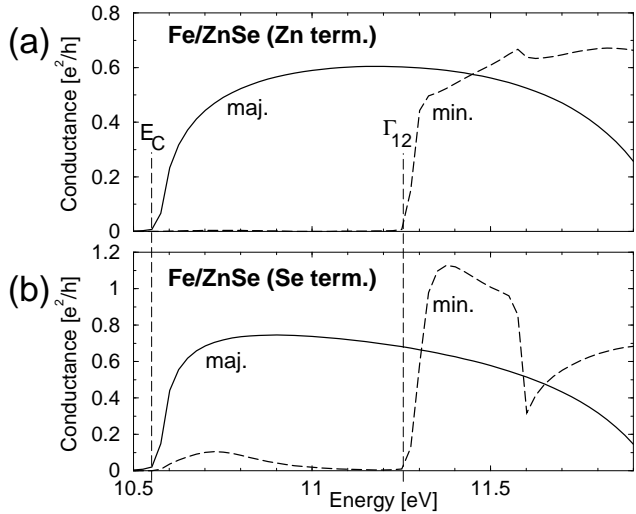,width=3.3in,angle=0}
    \caption{Injection of hot electrons from Fe into ZnSe with a Zn
      termination (a) 
      and a Se termination (b). For simplicity the conductance is
      calculated only at the $\Gbar$-point. The solid line shows the
      conductance in the majority and the dashed line in the minority
      band. The energy $E_C$ marks the
      bottom of the SC conduction band and $\Gamma_{12}$ refers to Fig.\
      \ref{fig1}.}
    \label{fig2}
  \end{center}
\end{figure}   
Firstly we discuss the injection process of hot electrons with Fe
states well above $E_F$. Although for hot spin injection states with
non-zero 
$\kp$ values also play a role, we consider here for simplicity only
states with normal incidence. The calculated transmission
probabilities for injection into ZnSe are shown in Fig.\ \ref{fig2}
for both spin directions, with Fig.\ \ref{fig2}(a) referring to a Zn
terminated interface and Fig.\ \ref{fig2}(b) to a Se terminated
one. 
The transmission starts at the energy $E_C$ of the SC conduction
band minimum. In the majority band the conductance strongly increases
to values of around 0.6 or 0.7 (in units of $e^2/h$), while the
conductance in the minority band is much smaller. As a result, the
spin polarization of the injected current is very large, for Zn
termination always larger than 97\%, for Se termination larger than
75\%. However, the situation completely changes, if the energy of the
injected Fe electrons exceeds the value $E_{\Gamma_{12}}$ of the
minimum of the minority $\Delta_1^\mathrm{Fe}$-band. There the
transmission in the minority band increases very sharply and even
overcomes the majority transmission, so that the spin polarization
changes sign. This clearly illustrates, that the absence of the 
$\Delta_1^\mathrm{Fe}$-state in the minority band leads for lower
energies to the very large spin polarization of the current. Similar
results are also obtained for the hot spin injection into GaAs(001),
resulting, for lower energies, even in polarizations extremely
close to 100\%.

The strong spin polarization can be understood from the different
spatial orientation and extent of the $\Delta_1^\mathrm{Fe}$ and
$\Delta_{2'}^\mathrm{Fe}$-states. The $\Delta_1^\mathrm{Fe}$-states
have $s$, $p_z$ and $d_{z^2}$ admixtures. In particular the $s$ and
$p_z$ 
components have large spatial extent and a strong overlap with the SC
states. Moreover the $d_{z^2}$ and in particular the $p_z$ orbitals
point directly into the SC, so that a large transmission is
possible. In contrast to this the minority
$\Delta_{2'}^\mathrm{Fe}$-states consist of in-plane $d_{xy}$ orbitals
which are much less extended and point in the wrong direction.

\begin{figure} [tb]
  \begin{center}
    \epsfig{file=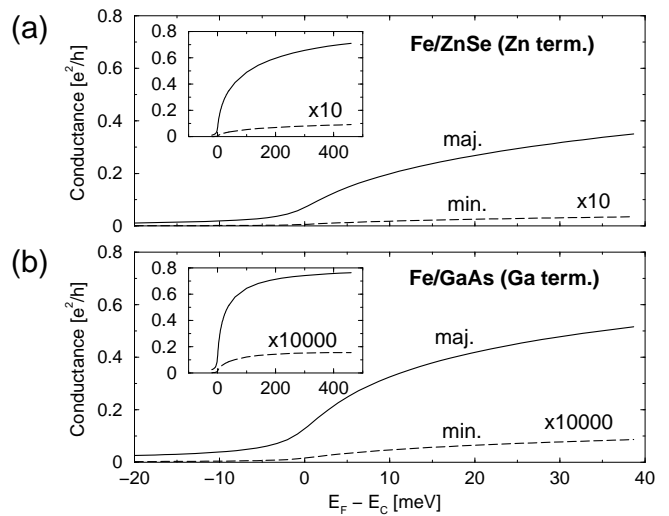,width=3.4in,angle=0}
    \caption{Energy dependence of the barrier-free injection of
      electrons at $E_F$ for a Fe$|$ZnSe junction with Zn termination
      (a) and a Fe$|$GaAs junction with Ga termination (b) at the
      $\Gbar$-point. The solid lines show the conductance in the
      majority and the dashed lines in the minority band. In (a)
      the minority conductance is enlarged by a factor of 10 and in
      (b) by a factor of $10^4$. The insets show the conductance in a
      wider energy range.}
    \label{fig3}
  \end{center}
\end{figure}   
To model the injection of electrons at $E_F$ we lower the potential in
the SC halfspace such that the Fermi level falls slightly above the
conduction band energy $E_C$. Here we consider two situations, by
simulating the injection process both without and with a tunneling
barrier. In the first case, we lower the potentials of the
3$^\mathrm{rd}$, 4$^\mathrm{th}$ and all further away SC ML by
the same constant value, so that $E_F - E_C$ becomes positive. We do
not change the potentials of the two SC ML closest to the
interface, since they are important for the interface
characteristics. By continuously varying the potential step, we
calculate then the conductance as a function of $E_F - E_C$. Fig.\
\ref{fig3} shows the resulting conductance at the
$\Gbar$-point for an Fe$|$ZnSe(001) junction with Zn interface
termination (Fig.\ \ref{fig3}(a)) and for an Fe$|$GaAs(001) junction
with Ga termination (Fig.\ \ref{fig3}(b)). 
The energy scale in the order of 10~meV refers to typical carrier
concentrations in a two-dimensional 
electron gas \cite{Grundler}. The insets show the results over a larger
energy region. The minority intensities are enhanced by a factor of 10
for ZnSe and by a factor of $10^4$ for GaAs. Thus the spin polarizations
are larger than 97\% for ZnSe and practically 100\% for GaAs. Very
similar results are also obtained for the other terminations not shown
here, i.e. the Se termination of ZnSe and the As termination of GaAs.

All the calculated results in Fig.\ \ref{fig2} and Fig.\ \ref{fig3}
suggest, that near the energy $E_C$ of the SC conduction band minimum,
the transmission probability varies for both spin directions as
$\sqrt{E_F - E_C}$ \cite{broadening}.
Since the square root-like behavior is the same for both the majority
and the minority electrons, the spin polarization remains constant for
$E \rightarrow E_C$. Moreover, in the interesting energy region of about
10~meV, the reduction of the conductance is rather modest. The
square-root like behavior of the transition probability can be
understood from a simple picture where a potential step in one
dimension is assumed. For a constant potential of height $V_B$ in the
right halfspace and a 
vanishing potential in the left halfspace, the transition probability
for an incident electron with energy $E = k^2$ into a transmitted
state with the same energy $E = V_B + k'^2$ is given by $T = \frac{4 k
  k'}{(k+k')^2} \cong 4 \frac{k'}{k} \propto \sqrt{E-V_B}$ for $k'
\rightarrow 0$.

\begin{table} [tb]
 \begin{ruledtabular}
 \begin{tabular} {lllll}
  Thickness $N$ & $P$ (Zn) & $P$ (Se) & $P$ (Ga)& $P$ (As) \\
  \hline
  8 ML & 96\% & 99.3\% & 99.99\% & 99.8\% \\
  32 ML & 86\% & 99.3\% & 99.99\%  & 99.6\%  \\
  80 ML & 80\% & 99.3\%  & 99.98\% & 98.6\% \\ 
  144 ML & 77\%  & 99.3\%  & 99.97\% & 97.6\% \\
  \hline
  32 ML (integr.) & 84\% & 96.9\% & 99.52\% & 99.4\%
 \end{tabular}
 \end{ruledtabular}
 \caption{Spin polarization of the current at the $\Gbar$-point for
  Fe$|$ZnSe and 
  Fe$|$GaAs systems with different tunneling  barrier thicknesses $N$. All four 
  terminations are shown: Zn
  and Se for a Fe$|$ZnSe junction and Ga and As for a
  Fe$|$GaAs junction. In the last row also the polarization is given
  for a 32 ML thick tunneling barrier
  when integrating over the whole two-dimensional Brillouin zone.}
 \label{table1}
\end{table}
To simulate the effect of a Schottky barrier, we modify the
above model by smearing out the potential step, i.e. by lowering the
external potential continuously over a distance of $N$
SC-ML. Within this barrier of $N$ ML thickness
effectively the Fermi level slowly increases with respect to the local
potential, from the ground-state value deep in the gap to an energy
value slightly above $E_C$. Assuming for this final position of $E_F$
a typical energy value $E_F - E_C$ = 10~meV, we list in Table\
\ref{table1} the resulting spin polarizations $P$ at the $\Gbar$-point
obtained for Fe$|$ZnSe(001) and Fe$|$GaAs(001) junctions with four
different barrier thicknesses of $N$ = 8, 32, 80 and 144 ML. As an
example for the polarization obtained by integrating over the
two-dimensional Brillouin zone, in the last row the polarization is
given for a 32 ML thick barrier. Since the integration affects both
spin channels approximately equally, the polarization is changed only
slightly. While in the case of Se, Ga and As termination the
polarization of the 
spin current is equally high ($\ge 97\%$) as in the barrier-free case,
we see a gradually lowering of the spin polarization for the Zn
termination, which however levels off at a value of about 77\% for
large barrier thicknesses. This effect arises from the existence of
minority interface states at the Fe$|$SC(001) interface. These states
of $\Delta_1$-symmetry lie within the 
$\Delta_1^\mathrm{Fe}$-bulk gap and become resonant due to the coupling with the
$\Delta_{2'}^\mathrm{Fe}$-band. In the case of Zn-termination, the interface state
at $\Gbar$ lies relatively close to $E_F$, i.e. 0.15eV below. Its
effect is to reduce the (positive) spin injection ratio. If this
state would coincide with $E_F$, its effect would be much bigger and
could even lead to a strong negative polarization. 

In summary, we have performed \textit{ab initio} calculations to
investigate the 
ballistic spin injection from a Fe half-crystal into ZnSe and
GaAs SCs. Three processes of injection have been considered:
the injection of hot electrons as well as the injection of electrons
at the Fermi level with and without an interface tunneling barrier. The
calculations demonstrate that the Fe$|$ZnSe and Fe$|$GaAs (001)
interfaces 
act as highly spin-polarizing filters yielding polarizations as
high as 99\%. This behavior can be traced back to some simple
properties of the band structure of Fe for normal incidence: the
majority states at the Fermi level have
$\Delta_1^\mathrm{Fe}$-symmetry and a strong $s$ and $p_z$ admixture,
so that they can  
couple well to the conduction band states of the SC, while
the Fe minority states at $E_F$ have a different symmetry and can
either couple only weakly or not at all to the SC states.
This picture becomes clearer, the more ordered the interface is, since
interface disorder breaks the $\kp$ conservation and can reduce the
spin polarization of the current.
Our calculations and the recent successful observation of a 2\% spin
injection in the Fe$|$GaAs(001) system \cite{Zhu} suggest, that much
larger spin injection efficiencies should be achievable.

\begin{acknowledgments}

The authors thank G. Schmidt for helpful discussions. This work was
supported by the RT Network \textit{Computational Magnetoelectronics}
(Contract RTN1-1999-00145) of the European Commission.

\end{acknowledgments}

% Bibliography -----------------------------------------


\begin{thebibliography}{99}
\bibitem{broadening}
 For numerical purposes a small but finite imaginary part of the energy
 is used in the evaluation of the Green's function, showing up in a
 Lorentzian broadening of the conductance in Fig.\ \ref{fig3} around $E_C$.
\end{thebibliography}
\end{document}